# Features of nanocontact formed using point electrical breakdown of a niobium oxide nanolayer

S.I. Bondarenko, A.V. Krevsun, V.P. Koverya

B. Verkin Institute for Low Temperature Physics and Engineering of the National Academy of Sciences of Ukraine, Kharkiv 61103, Ukraine
E-mail: bondarenko@ilt.kharkov.ua

The electrical characteristics of a point section of niobium oxide between a massive cathode made of a superconducting indium-tin alloy and a niobium film before the oxide breakdown and the nanocontact that appears after the breakdown of the oxide at room temperature have been studied. The dependences of the point breakdown voltage of the oxide and the resistance of nanocontacts on the method of breakdown and the thickness of the oxide in the range of 15-60 nm have been established. In particular, a minimum value of the breakdown current has been established (about 1 µA), below which breakdown does not occur. The voltage-current characteristic (VCC) of the oxide has a semiconductor character and indicates the dependence of the breakdown voltage of the oxide on its polarity. The VCC of the nanocontact is nonlinear, which is associated with phase transitions in the nanocontact material caused by its heating by the transport current.

## 1. Introduction

Point electrical contacts (PECs) have been studied for more than 100 years [1,2]. Most of the works are devoted to contacts between non-superconducting metals. In the 60s of the last century, the study and application of contacts made of superconducting metals began [3,4]. Interest in the study of superconducting PECs is caused by the fact that they have the properties of Josephson contacts [5]. Such contacts have found application as detectors and generators of microwave radiation and superconducting quantum interferometers (SQUIDs) [6]. The most widely used superconducting PECs are of two types: with mechanical pressing of electrodes [3,4] and those obtained by mechanical puncture of a thin layer of insulator between film electrodes [7, 8, 9]. The latest PECs were manufactured in two versions. In the first of them [7,8], after piercing the film electrodes and the insulator between them with a thin steel needle, an additional stabilizing film from the material of the upper electrode was sputtered onto the puncture site, and in the second version [9] there was no additional film. The advantages of these technologies compared to the technology for creating traditional Josephson film junctions [10] were the ease of their manufacture. But in this case, the formation of a pressing point contact with the required normal resistance requires the creation of sufficiently complex mechanical control and adjustment of the process of bringing the electrodes together, repeated after each cycle of cooling and heating of the contact**.** Because of this, the possibility of creating compact and reliable systems from such contacts is practically excluded. Mechanical puncture contacts do not have this drawback, but have a significant spread in resistance values due to a complex, irreproducible microstructure in the form of several parallel-connected microcontacts with different resistances. The latter circumstance complicates theoretical calculations of the properties of the contacts and worsens the electromagnetic properties of PECs and SQUIDs based on them. We were the first to develop [11,12,13] Josephson contacts (JCs) and quantum interferometers using electrical breakdown of a dielectric nanolayer in the form of niobium oxide with a precisely specified thickness between film electrodes of niobium (bottom electrode) and lead or indium-tin alloy ( top electrode).

The choice of these materials with a critical temperature above 4.2 K is an important advantage, because allows quantum devices to operate in liquid helium without pumping out helium vapor. The breakdown of other dielectrics between tin films was also studied by I.K. Yanson [14] with the aim of using the resulting contacts for microcontact spectroscopy of various materials. The method of manufacturing JCs using dielectric breakdown between film electrodes allows one to obtain a single metal contact between superconducting films [13, 14]. In fact, this is a new type of PEC, resulting from local heating of a dielectric by current and subsequent explosive melting of one or both electrodes, followed by their microwelding. In this case, the length of the contact in the form of a metal nanobridge between the electrodes is determined by the thickness of the oxide being broken through and after a low-temperature superconducting transition can be equal to the coherence length of of the catode.

Previously [15, 16], we studied a contact in the form of a metal nanobridge between two film electrodes, which appears after the electrical breakdown of a niobium oxide nanolayer with a thickness of 30 nm. In this work, we studied the nanocontact that appears after a point breakdown of niobium oxide with a thickness of 15, 30, 45, 60 nm between a massive cathode pressed to the oxide on a quasi-point area of about 6 μm$^2$ and a film anode. The thickness of such oxides is tens of times greater than the thickness of tarnish oxides on metals, the breakdown of which, called fritting [2, 3], has been studied quite well, in contrast to the breakdown of "thick" oxides that we are studying. Before carrying out these studies, unresolved issues were: the electrical properties of the point section of niobium oxide before breakdown, the dependence of the point breakdown voltage of "thick" niobium oxides between the massive cathode and the film anode and the resistance of the bridges on the thickness of the oxide, as well as the transport properties of such bridges.

The purpose of the experimental studies presented in this article is to determine the formation features and properties of a new type of PEC formed as a result of a point electrical breakdown of a niobium oxide nanolayer with a thickness of 15 nm to 60 nm between film and massive electrodes made of superconducting materials.

**2. Setting up experiments.**

Figure 1 shows a diagram of the experimental structure. It consists of an anode in the form of a niobium (Nb) film on a silicon (Si) substrate, a thin layer of niobium oxide on the surface of the film, and a spring-loaded massive cathode (in the form of a u-shaped copper conductor coated with an indium-tin alloy) in contact with the oxide surface. The structure has electrical connections through a limiting resistance ($R_L$) to an electrical voltage source ($U_{in}$) and a device (not shown) for measuring voltage ($U_{ox}$) on the oxide.

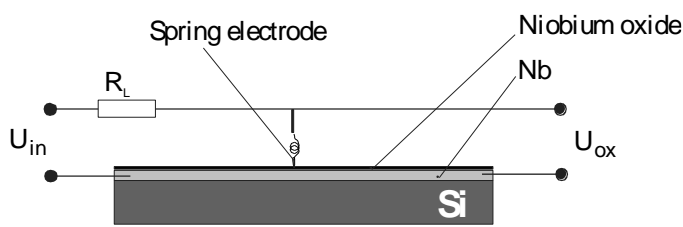

Fig. 1 Schematic of the experimental structure.

The mechanical and at the same time electrical connection of the cathode with the oxidized anode is called point only conditionally by analogy with clamping contacts without an intermediate oxide, used in weak superconductivity [3,4]. In reality, any such contact differs from a mathematical point in that it has a very specific small area, which can be difficult to measure or calculate. In this work, an attempt is made to determine this area indirectly (see Section 3).

To ensure mechanical contact of the cathode with the oxide before breakdown, a device described in [15] was used. The dimensions of the anodic niobium film and the method of oxide formation on it are also given there.

Figure 2 shows a complete electrical diagram of a device for applying electrical voltage between the cathode and anode with "point" contact of the cathode with the oxide surface. The circuit is a further development of the breakdown control circuits we previously used [15, 16] and allows for breakdown in three ways:
- a slow increase in direct voltage (SIV) on the oxide in the range of 0-30 V until breakdown,
- a slow increase in single-cycle alternating voltage,
- a method of the combined action (MCA) of direct voltage and capacitor discharge [16].

A generator TEC 9 was used as a source of adjustable direct voltage, and a G3-18 generator was used as a source of alternating voltage with a frequency from 60 Hz to 16 kHz. The DC voltage on the bridge was recorded using a digital voltmeter V7-28, and the alternating voltage – using an oscilloscope S1-83.

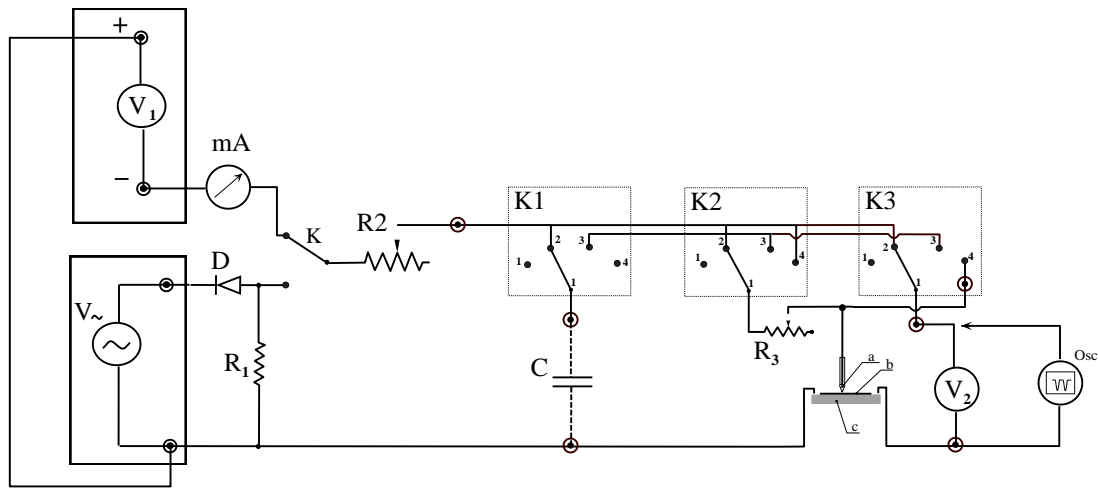

Fig. 2 Electrical circuit of a device for applying voltage to an experimental structure consisting of a cathode (a), oxide (b) and anode (c) in one of three ways: from a constant voltage source $V_1$ without a capacitor C, with a capacitor C, or from an alternating voltage source $V_\sim$.

To switch voltage sources, a key K and a drum-type switch with three groups of contacts K1, K2, K3 were used. The voltage on the oxide was measured either with a digital voltmeter $V_2$ or an oscilloscope (Osc). Single-cycle rectification of alternating voltage was performed using a diode (D) type.... Limiting $R_L = R_2$ and discharge resistance $R_3$ were adjustable. The resistance of $R_1$ was 1 kOhm. The current through the structure was measured with an ammeter (mA) type M253.

The work consists of three parts:
- determination of the area of "point" contact of the cathode with the surface of the oxide before breakdown and its electrical properties,
- study of the dependence of the breakdown voltage of the oxide and the electrical resistance of the metal nano-sized bridge that appears inside it,
- study of the transport properties of the formed bridges.

The characteristic electrical properties of the oxide are presented using the example of the voltage – current characteristic (VCC) and the electrical capacitance of a point contact of a massive cathode with a 30 nm oxide before its breakdown. The capacitance was measured using digital set

E7-8. The dependences of the breakdown voltage and bridge resistance on the oxide thickness were obtained for oxide thicknesses of 15, 30, 45 and 60 nm. The current-voltage characteristic of a bridge across an oxide with a thickness of 30 nm gives an idea of its transport properties.

All measurements were performed at room temperature (300K).

**3. Results of experimental studies and their discussion.**

**3.1 Electrical capacity and resistance of the oxide at the point where the cathode is located before breakdown**.

Pressing a cathode coated with a relatively soft indium-tin alloy against a niobium oxide surface can lead to the formation of a flat area in the area where it touches. In the experiments performed, the force of action of the cathode on the oxide was about $10^{-4}$ N, which corresponds to a load of 1 g. With such an impact on the alloy, with a contact area of several square microns, a mechanical stress arises in it that exceeds Young's modulus for metals such as tin and aluminum [17] and, even more so, can exceed the elastic limit for a softer indium-tin alloy. In this regard, the electrical contact between the cathode and anode before breakdown can be represented in the form of a small-size flat capacitor with a dielectric in the form of niobium oxide, which has a capacitance ($C_p$) and a leakage resistance connected in parallel. The value of the capacitance $C_p$ depends on its area ($S$) and the relative dielectric constant of the oxide ($\varepsilon$) [18]:

$$C_p = \varepsilon_0\, \varepsilon\, S/s, \qquad (1)$$

where $\varepsilon_0$ and $s$ are, respectively, the absolute dielectric constant and the oxide thickness. Formula (1) allows you to calculate the area of mechanical quasi-point contact with the oxide if the capacitance and relative dielectric constant $\varepsilon$ are known. From our measurements it follows that at $s = 30$ nm, $C_p = 0.2\times10^{-12}$ F and $\varepsilon \approx 40$ [19] we have $S \approx 6\times10^{-12}$ m$^2$, that is, approximately an area equal to $2.5\times2.5$ μm$^2$.

To determine the resistance of niobium oxide at a temperature $T = 300$K before its breakdown, the voltage – current characteristic (VCC) of a point section of the oxide with a thickness of 30 nm was measured (Fig. 3).

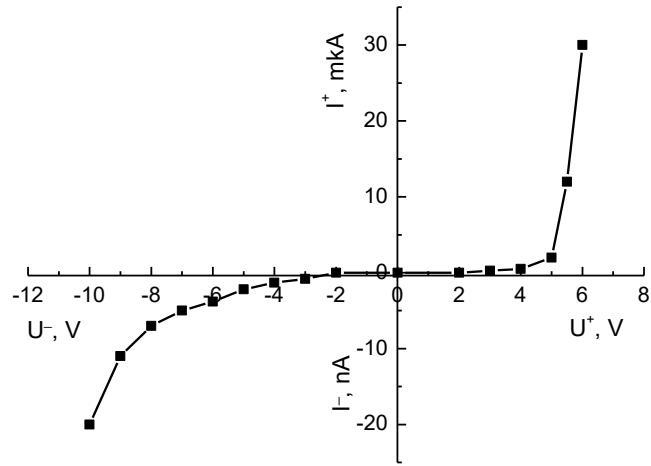

Fig. 3 Voltage-current characteristic of a point section of niobium oxide before breakdown.

On Fig. 3 it can be seen that the dependence of current on voltage for its two polarities $U^-$ and $U^+$, i.e. with currents of different directions, sharply asymmetrical. Oxide breakdown occurs at 10 and 6 volts, respectively. In this case, the absolute value of the current in different directions of its flow differs by almost two orders of magnitude. The lowest current corresponds to a negative potential at the cathode. Thus, a point region of niobium oxide has semiconductor properties before its breakdown. In accordance with Fig. 3, the electrical resistance of a point section of the oxide is a nonlinear function of voltage and is different for the negative and positive branches of the dependence of current on voltage (Fig. 4 and Fig. 5).

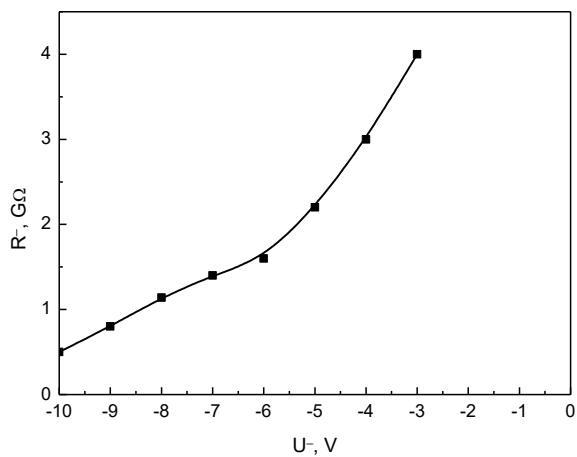

Fig. 4 Dependence of the resistance of a point section of niobium oxide on voltage at a negative potential at the cathode.

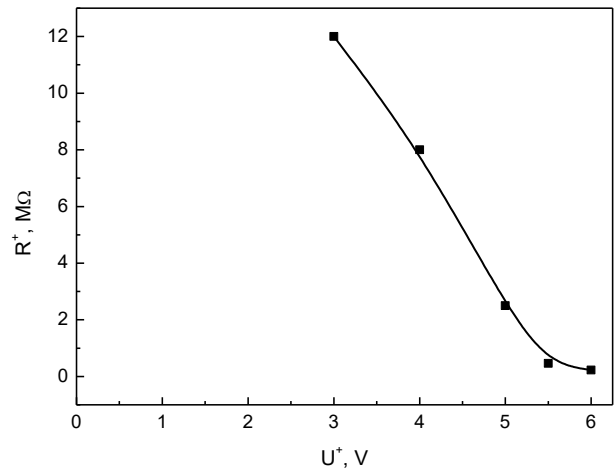

Fig. 5 Dependence of the resistance of a point section of niobium oxide on voltage at a positive potential at the cathode.

After breakdown of the oxide and opening of the mechanical contact of the massive cathode with the surface of the oxide, a microscopic metal inclusion can be seen at the site of breakdown. This means that during breakdown, part of the cathode coating of the indium-tin alloy was melted by the breakdown current and remained on the oxide surface. A photograph of the oxide surface after its breakdown in several areas after lifting the movable cathode is shown in Fig. 6.

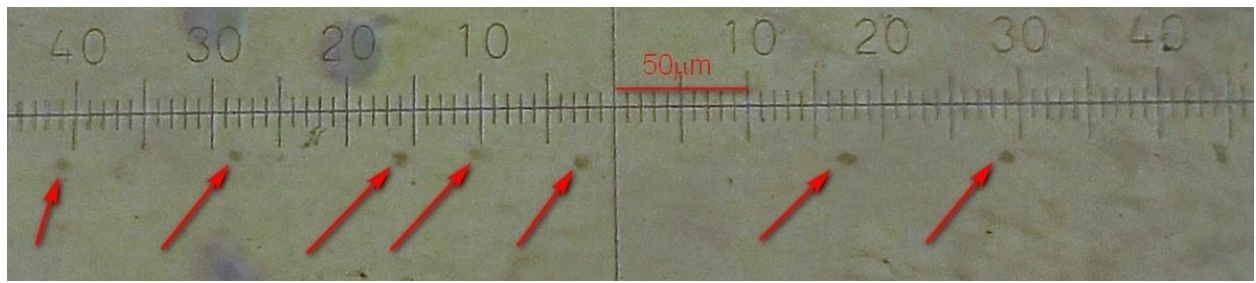

Fig. 6 Microphotograph of a section of niobium oxide with traces of seven breakdown sites on its surface.

The size of the dark dots in the photograph is about 5 microns. This size is close to the size of the mechanical contact of the massive cathode with the oxide, obtained by measuring its electrical capacitance using formula (1). The obtained results confirm the assumption of loss of elasticity of the coating at the point of mechanical contact before breakdown and melting of the cathode coating during breakdown. In addition, the size of the contact area of the massive cathode with the surface of niobium oxide has been established. Thus, the microstructure of the "point" region of electrical contact between the cathode and anode after breakdown is represented by series-connected sections in the form of a cathode coating on the surface of the oxide and a nano-sized bridge [15] of the coating material inside the oxide.

**3.2 Study of the dependence of the point breakdown voltage of the oxide and the electrical resistance of the resulting bridges on the thickness of the oxide, the breakdown method and the electrical parameters of the breakdown circuit.**

**3.2.1 Dependence of oxide breakdown voltage ($U_b$) on its thickness for various breakdown methods.**

Figure 7 and Fig. 8 show the experimental dependences of the point breakdown voltage $U_b$ of niobium oxides on their thickness.

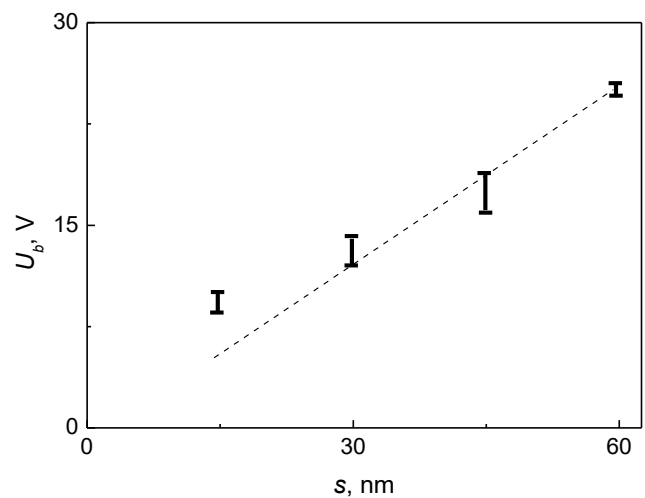

Fig. 7 Dependence of oxide breakdown voltage ($U_b$) on its thickness ($s$) when using SIV. The vertical lines correspond to the spread of $U_b$ values at four different points on the oxide surface.

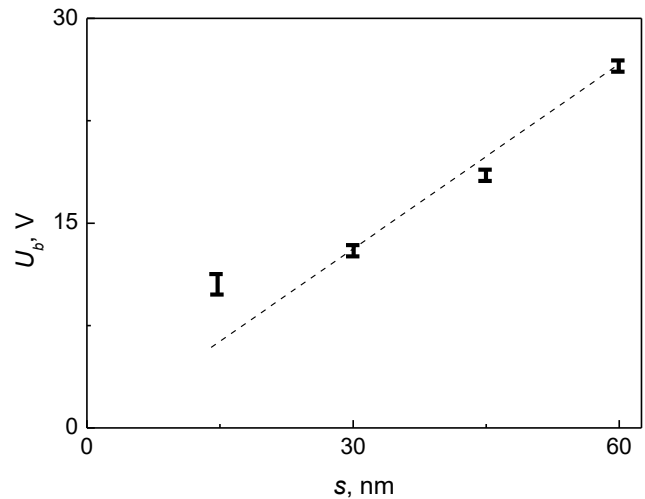

Fig. 8 Dependence of oxide breakdown voltage ($U_b$) on its thickness when using MCA. The vertical lines correspond to the spread of $U_b$ values at four different points on the oxide surface.

Figure 7 shows the breakdown voltage $U_b$ of niobium oxide at four different points on the surface of niobium oxide for each of its four thicknesses (15, 30, 45, 60 nm) with a slow increasing the voltage (SIV) across it. In this case, the value of the limiting resistance $R_L = R_2$ in the breakdown device was equal to 1 kΩ. It can be seen that $U_b$ of oxides has a noticeable spread in absolute values (up to 30%) for oxides with a thickness of 30-60 nm and less (10-15%) for an oxide with a thickness of 15 nm. The scatter can be explained by the influence of local (within the area of point contact of the cathode with the oxide) heterogeneity of niobium oxide when using the SIV method. In addition, two features are visible in the $U_b(s)$ dependence. One represents the linear region for $s$ = 30-60 nm, and the second represents the deviation from linearity at $s$ = 15 nm. Accordingly, in the linear section the breakdown electric field strength $E_b = U_b/s = 4.3 \times 10^6$ V/cm does not change, but at s = 15 nm it increases and amounts to $E_b \approx 6 \times 10^6$ V/cm. The increase may be explained by better uniformity of the finer niobium oxide.

Figure 8 shows the dependence of the point breakdown voltage of the oxide on the thickness of the oxide when using the combined electrical breakdown (MCA). In this case, a capacitor with a capacity of 180 μF was used, and the limiting and discharge resistances were equal to $R_2$ = 1 kΩ and $R_3$ = 10 Ω, respectively. It can be seen that the breakdown voltage of the oxides has a significantly smaller spread of values at different points on the surface of the oxides in comparison with the SIV method (Fig. 7). This can be explained by the difference in the process of formation of oxide breakdown in the SIV and MCA methods. In this case, in the $U_b(s)$ dependence, just as for SIV, there is a deviation of the breakdown voltage from the linear dependence at an oxide thickness of 15 nm.

## 3.2.2. Dependence of the breakdown voltage of oxides of various thicknesses (*s*) by the MCA method on the capacitance value (*C*) of the capacitor.

Table 1 show the breakdown voltage values of the MCA for oxides of various thicknesses depending on the capacitance value of the capacitor.

Table 1. Breakdown voltage ($U_b$) of oxides with a discharge resistance $R_p=R_3=10$ Ω and a limiting resistance $R_L = R_2 = 1$ kΩ at different capacitor capacities.

| | $U_b$ (V) | | | |
|---|---|---|---|---|
| *s* / *C* | 0.96 µF | 0.16 µF | 0.015 µF | $10^{-3}$ µF |
| 15 | 9.2-9.7 | 8-9 | 8.5-8.8 | 9.5-10 |
| 30 | 17-20 | 20-25 | 20-27 | 25 + N.B. before 30V |
| 45 | 20-30 | 20-30 | - | N.B. before 30V |
| 60 | 20-30 | 25-30 | - | N.B. before 30V (N.B. – No Breakdown) |

The data presented in Table 1 allows us to see three of their features:

- at $C = 10^{-3}$ µF and *s* > 15 nm, breakdown either does not occur at all at voltages up to 30 V, or occurs only at some points of the oxide. In this case, an oxide with a thickness of 15 nm breaks through, but the breakdown voltage becomes greater than with large capacitors,
- breakdown voltage at oxide thicknesses of 30-60 nm almost does not depend on the capacitance value of the capacitor if $C = 0.96$ µF and 0.16 µF, and is noticeably higher than in the SIV mode at the same value of $R_2$,
- the breakdown voltage of an oxide with a thickness of 15 nm is practically independent of the capacitance of the capacitor in the range from 0.96 µF to 0.015 µF.

The first feature can be explained by the insufficient energy stored in the capacitor with a capacity of $10^{-3}$ µF. As estimates show, this energy may be less than the energy required to maintain the avalanche-like process [20] of evaporation of the nanobridge material that occurs during the breakdown of an oxide with a thickness of more than 15 nm.

The second feature that distinguishes a point breakdown using a capacitor from an SIV breakdown between planar film electrodes is explained by the fact that in the latter case a breakdown of an electrically weakened dielectric occurs. The reason for this is the penetration (during deposition of a film cathode) of film atoms into the surface inhomogeneities of the oxide. As a result of these micro-inhomogeneities in the relief of the cathode film, the electric field strength between the electrodes required for breakdown may decrease. The consequence of this is a decreasing of breakdown voltage.

The third feature is probably the breakdown property of niobium oxide with such a minimum thickness for our experiments (15 nm).

The absence or weak influence of the capacitance value and energy stored in the capacitor on the breakdown voltage value can be useful in the process of applied developments using such point superconducting contacts.

### 3.2.3 Dependence of the breakdown voltage of oxides of various thicknesses when using SIV on the value of the limiting resistance.

Table 2 shows the magnitude of the point breakdown voltage of niobium oxides of various thicknesses using SIV, depending on the value of the limiting resistance in the breakdown circuit.

Table 2. Breakdown voltage ($U_b$) of oxides of different thicknesses ($s$) at different values of the limiting resistance ($R_L=R_2$).

|  | $U_b$ (V) | | | |
|---|---|---|---|---|
| $s / R_L$ | $10^5$ Ω | $10^6$ Ω | $3\times10^6$ Ω | $10^7$ Ω |
| 15 | 10-12 | 6.8-8.4 | 7-10 | 5+N.B. before 30V |
| 30 | 13-18 | 12.5 -13.5 | 14 +Voltage jumps | 12 + Voltage jumps |
| 45 | 17-22 | 17-20 | - | 9 + Voltage jumps |
| 60 | 23-27 | 24-27 | - | N.B. before 30V (N.B.- No Breakdown) |

An important feature of point breakdowns with large limiting resistances is the low breakdown current. As the resistance increases from $10^5$ to $10^7$ Ω, it decreases for different oxide thicknesses from 100-200 µA to 1-2 µA. As it turned out, this feature determines the entire range of possible values of the breakdown voltage of the niobium oxide thicknesses we studied. From the table it follows:
- with a breakdown current of less than 2 µA, the breakdown becomes either unstable or does not occur at all up to 30 V,
- with a breakdown current from 100 µA to 20 µA, the breakdown voltage decreases with decreasing oxide thickness and has a spread of its values by 10-15% at each thickness. With an oxide thickness of

15 nm, the breakdown voltage is disproportionately lower compared to oxides with thicknesses of 30, 45 and 60 nm,

- average breakdown voltage values at a breakdown current from 100 µA to 20 µA are slightly higher than the breakdown voltage values of oxides between film electrodes [15,16].

The first feature of breakdown at low values of its current can be explained by the low density of this current, because according to the theory of electron breakdown [20], for the formation of an electron avalanche it is necessary to have a current density of $10^9$ A/cm$^2$. According to estimates, at a breakdown current of 1 µA, its density in a nanobridge with a diameter of 10 nm can be only $10^6$ A/cm$^2$.

The second feature of such a breakdown can be explained by the better homogeneity of the oxide with a thickness of 15 nm compared to thicknesses of 30-60 nm.

The third feature of the increased average breakdown voltage (in comparison with the electrical strength of the oxide between the film electrodes) can be explained by the higher electrical strength of the oxide point section under the clamping cathode.

### 3.3. Dependences of the resistance of bridges arising during breakdown of the oxide on its thickness and method of breakdown.

Tables 3, 4, 5 and 6 present the resistance values of bridges formed using point electrical breakdown of niobium oxide of various thicknesses using different breakdown methods.

Table 3. Bridge resistance ($R_b$) during breakdown of oxide of various thicknesses ($s$) using SIV at $R_L=R_2 = 1$ kΩ.

| $s$ (nm) | $R_b$ (Ω) |
|---|---|
| 15 | 9-10.5 |
| 30 | 3.4-8.8 |
| 45 | 5.6-13 |
| 60 | 5.8-8 |

Table 4. Bridge resistance ($R_b$) during breakdown of oxide of various thicknesses using a MCA at $C = 180$ µF, $R_3 = 10$ Ω, $R_L = R_2 = 1$ kΩ.

| $s$ (nm) | $R_b$ (Ω) |
|---|---|
| 15 | 3.7-6.4 |
| 30 | 3.2 -4.2 |
| 45 | 3.8-5.8 |
| 60 | 3.0 -3.6 |

Table 5. Bridge resistance ($R_b$) during breakdown of oxide of various thicknesses using a MCA depending on the capacitance value of the capacitor $C$ (µF) at $R_3=10$ Ω, $R_L = R_2 = 1$ kΩ.

| | $R_b$ (Ω) | | | |
|---|---|---|---|---|
| $s / C$ | 0.96 µF | 0.16 µF | 0.015 µF | $10^{-3}$ µF |
| 15 | 4-5 | 9.9-12.8 | 7.8-12.2 | 6.9-8.2 |
| 30 | 3.4-4 | 3.5-4.2 | 4-6 | 14.8+ N.B. before 30V |
| 45 | 3.6-4.2 | 3.8-4.3 | - | N.B. before 30V |
| 60 | 4.8-5.3 | 10-12 | - | N.B. before 30V |

Table 6. Bridge resistance ($R_b$) during breakdown of oxide of various thicknesses using SIV depending on the value of the limiting resistance ($R_L=R_2$).

| | $R_b$ (Ω) | | | |
|---|---|---|---|---|
| $s / R_L$ | $10^5$ Ω | $10^6$ Ω | $3\times10^6$ Ω | $10^7$ Ω |
| 15 | 6-12 | 3-10 | 5-20 | 20- No Breakdown before 30V |
| 30 | 5-11.5 | 11.3-18 | Resistance jumps | Resistance jumps |
| 45 | 23-960 | 16-32 | - | Resistance jumps |
| 60 | 13-60 | 5.8 – 52 | - | No Breakdown before 30V |

From the tables it is clear that:

- the spread in the resistance values of bridges formed by a point breakdown using SIV is two or more times greater (excluding bridges at $s = 15$ nm and $R_2 = 1$ kΩ), exceeds the spread in the breakdown voltage values. This is especially evident at $R_2 = 10^5 – 10^7$ Ω,

- resistance of bridges with a thickness of 15 nm obtained by the SIV method at $R_2= 1$ kΩ, exceeds the resistance of bridges with thicknesses of 30-60 nm. Thus, in the $R_b(s)$ dependence, a minimum of $R_b$ values is formed in the thickness region of 30 nm,

- resistance of bridges obtained by point breakdown using MCA, less resistance of bridges obtained using SIV. In this case, the resistance value of these bridges decreases with increasing capacitance $C$.

The first of these properties of the bridge resistance can be explained by the fact that its value depends not only on the breakdown voltage. The resistance is also affected by the breakdown current density and the microstructure of the oxide at the breakdown point. These parameters may differ at different points of the oxide.

The second property may be a consequence of bridge diameter decreasing due to the decreasing of its length, equal to the oxide thickness, to 15 nm.

The third property can be explained by the fact that in the MCA method the bridge is formed mainly due to the energy of the pulsed discharge of the capacitor. As the capacitance of the capacitor increases, the energy entering the breakdown region increases, which probably contributes to more complete ionization of the electrically conductive channel and the formation of a well-conducting metal bridge after cooling.

**3.4 Oxide breakdown voltage and bridge resistance when using an alternating voltage generator.**

Measurements of the breakdown voltage amplitudes and the resistance of the bridges formed during breakdown were carried out at a unipolar rectified alternating voltage at frequencies of 67, 376, 480, 5000 and 10000 Hz. In all experiments, a negative electrical potential was present at the massive catode. The oxide thickness was 15, 30, 45 and 60 nm, the limiting resistance varied from 1 k$\Omega$ to 100 k$\Omega$ A comparison of the breakdown voltage and bridge resistance values obtained under the influence of alternating voltage and direct voltage showed that the magnitude and spread of their values are close to each other and do not depend on frequency. In these experiments, one of the oscilloscope beams corresponded to the rectified voltage of the generator, and the other to either the voltage across the oxide or the voltage across the limiting resistance. The latter voltage was proportional to the current through the oxide (Fig. 9).

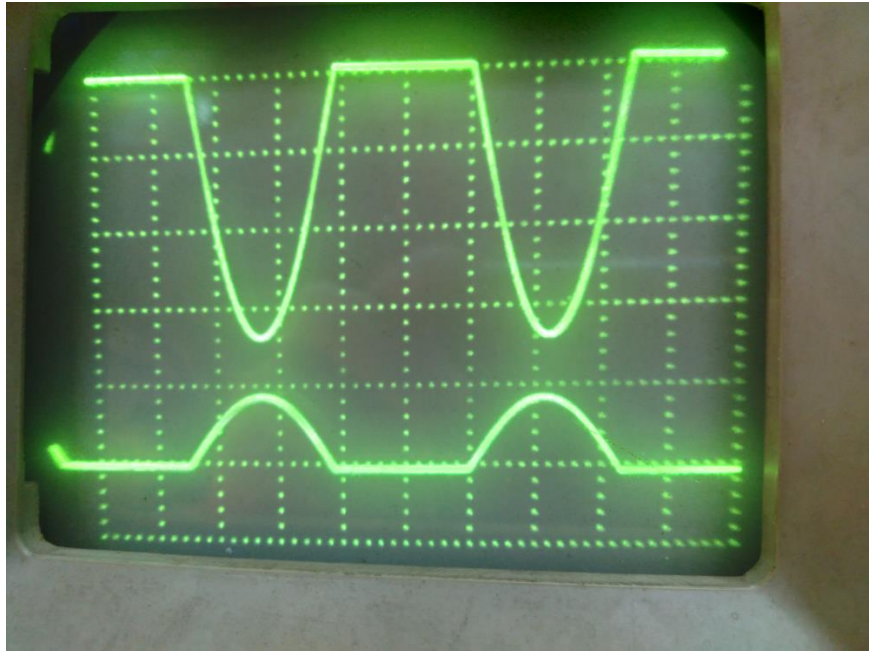

Fig. 9. Image on the oscilloscope screen of a single-cycle voltage across a limiting resistance with a frequency of 480 Hz (upper beam) and the voltage on the bridge after breakdown of niobium oxide with a thickness of 30 nm (lower beam). In one cell horizontally there is about 0.5 ms, in one cell vertically for the upper beam there is 5 V, and in one cell for the lower beam there is 200 mV.

**4. Study of the transport properties of point nanocontacts.**

The transport properties of various contacts are usually determined by their voltage – current characteristic s (VCCs). The most well-known in works with superconducting contacts are the VCCs at various cryogenic temperatures. Much less frequently mentioned are the VCCs of contacts between electrodes made of superconducting materials, measured at room (laboratory) temperature. At the same time, such current-voltage characteristics can provide important evaluation information about the potential applications of these contacts at low temperatures, as well as about their critical states at room temperature. This is due to the fact that in the region of weak superconductivity, experimenters are forced to perform auxiliary measurements, in particular, with film structures at room temperature. In this case, for example, in Dayem film bridges, a measuring current of 10 mA for the VCC can have a density about $10^5$ A/cm$^2$.

In some cases, such a current can heat up and melt the bridge, since it is known that fuses in electrical networks are designed to blow at a current density of $10^4$-$10^5$ A/cm$^2$. In addition, with the current state of rapidly developing electronics and electrical engineering, random electromagnetic interference exists in the surrounding space and in the wired network, which can act on structures with Josephson contacts and destroy them even at room temperatures. Therefore, it is necessary to know the transport capabilities of weak contacts. For this purpose, we carried out for the first time at an ambient temperature of 300 K (t = $27^0$ C) measurements of the VCCs of Josephson bridges manufactured by the method of electrical breakdown of a dielectric nanolayer in the form of niobium oxide with a thickness of 15 nm. Figure 10 shows the VCC of one of these bridges. A direct current was used in the range from zero to a critical value (33mA), upon reaching which the voltage on the bridge increases abruptly. In addition, at a current of 15 mA a kink appears on the VCC, which is clearly visible due to the mismatch of the line continuation of its initial section with the experimental part of the VCC.

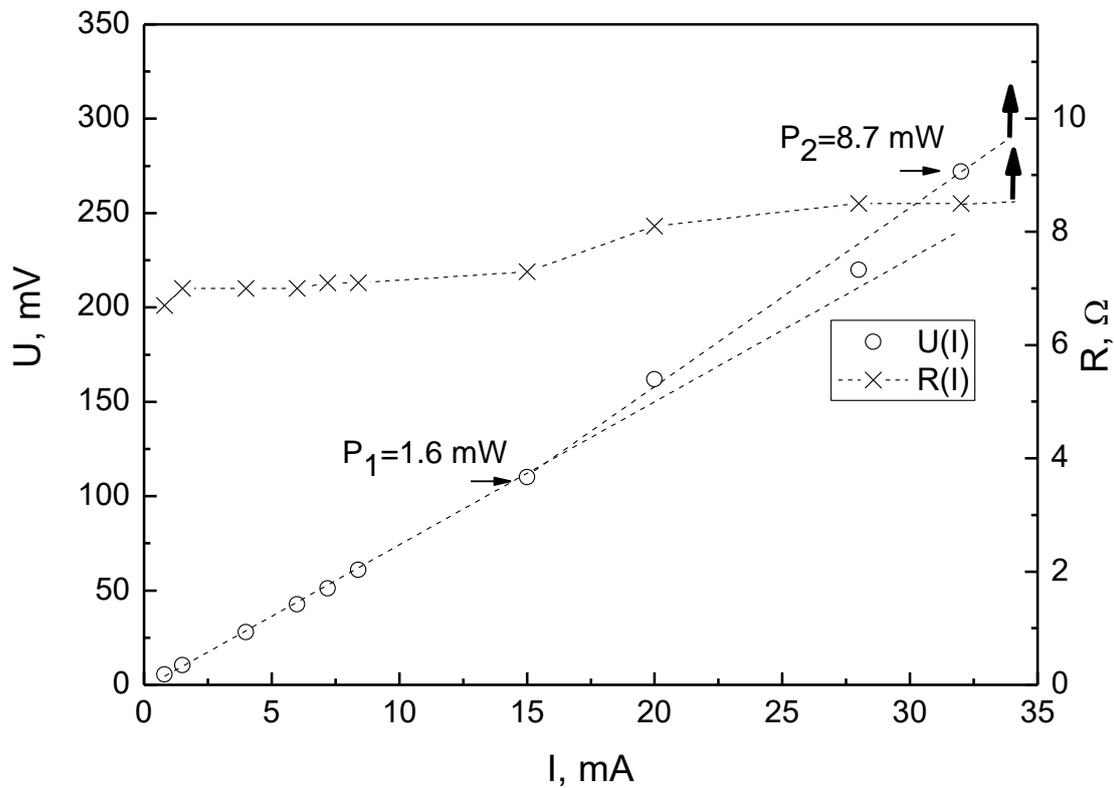

Fig. 10. Voltage- current characteristic $U(I)$ of the bridge formed by electrical breakdown of niobium oxide with a thickness of 15 nm and the dependence of the bridge resistance on the current $R(I)$. The horizontal arrows show the values of the thermal power $P_1$ and $P_2$ released in the bridge at currents of 15 and 32 mA, the vertical arrow shows the place on the voltage- current characteristic where a sharp increase in voltage occurs at a current of about 33 mA.

The second graph in Fig. 10 is the dependence of the bridge resistance on the current $R(I)$. Resistance was determined as the quotient of the voltage divided by the current through the bridge. The flow of current through the bridge causes it to heat up. The dependence of the bridge resistance ($R_t$) on its temperature ($t$) is described by the formula [18]:

$$R_t = R_{20} + R_{20}\, \alpha\, (t\text{-}20),  \qquad (2)$$

where $R_{20}$ and α are, respectively, the bridge resistance at a temperature of 20 $^0$C (i.e. at the initial voltage values on the current-voltage characteristic) and the temperature coefficient of the bridge resistance. At currents of 0-15 mA, the resistance increases slightly compared to $R_{20} = 6.9$ Ω. At currents of 15 -20 mA, the rate of resistance increasing increases. With a resistance of $R = 8.5$ Ω, the calculated temperature of the bridge (α=2.3×10$^{-3}$ /C$^0$) is about 120$^0$C. This temperature is close to the melting point (126$^0$ C [22]) of the indium-tin alloy. Consequently, an anomalous increase in the resistance of the bridge at currents of 15-20 mA occurs due to the phase transition of the solid metal of the bridge into the liquid state. At currents of 20-32.5 mA, the rate of resistance growth decreases. A "plateau" forms on the $R(I)$ dependence. The bridge is in a liquid state. At a current of about 33mA, there is a multiple sharp increase in resistance. It can be assumed that this is caused by the onset of boiling of the indium-tin alloy due to heating by the transport current.

One more feature of the formed bridge should also be noted. The bridge was formed at a breakdown current of 0.14 mA with a calculated density of about 0.4×10$^8$ A/cm$^2$. When the current increases to 15 mA, the resistance of the bridge changes little, but the current density in the stable state of the bridge increases by two orders of magnitude (up to 0.4×10$^{10}$ A/cm$^2$). This may indicate the increased resistance of such a bridge to current overloads that arise in practice.

**Conclusions.**

As a result of experimental studies of point contact between a massive and film electrode through a nanolayer of niobium oxide, the semiconducting properties of an oxide with a thickness of 15 nm and an area of about 6 square microns were established. At negative and positive potentials on a massive electrode made of an indium-tin alloy, electrical breakdown of the oxide occurs at a voltage of 10 and 6 volts, respectively.

It has been established that the breakdown electric field strength is about 4×10$^6$ V/cm for nanooxides with a thickness of 30, 45, 60 nm, and with a thickness of 15 nm it increases to 6×10$^6$ V/cm.

The resistance of bridges arising after the breakdown of an oxide with a thickness of 15 nm is greater than after the breakdown of oxides with a thickness of 30, 45 and 60 nm.

The magnitude of the breakdown voltage of oxides of different thicknesses and, especially, the magnitude of the resistance of the bridges that arise after the breakdown have scatter of values, which can be caused by point heterogeneity of the oxide structure.

A study of the features of a point breakdown depending on the value of the limiting resistance that determines the breakdown current in the SIV mode showed that at breakdown currents less than 1-2 µA, breakdown either does not occur or occurs only in some areas of the oxide.

At breakdown currents of about 10 mA, the magnitude and spread of resistance values of point bridges is greater than that of bridges formed between film electrodes.

A study of the characteristics of breakdown depending on the capacitance value of the capacitor in the breakdown circuit in the MCA mode showed that with a capacitance of less than $10^{-3}$ µF breakdown does not occur. The cessation of breakdown can be explained by the insufficient magnitude and duration of current flow through the oxide, necessary for the formation of an avalanche of electrons, characteristic of the phenomenon of point breakdown.

The transport properties of the bridge formed as a result of the breakdown of niobium oxide with a thickness of 15 nm are manifested in its nonlinear current-voltage characteristic and the unusual dependence of the bridge resistance on current. It is shown that their appearance is explained by a change in the phase state of the bridge as the current through it increases. At low currents (up to 15 mA) the bridge is in a solid state, at currents of 15-20 mA the bridge begins to melt, and at currents of 20-32 mA it is in a liquid state. At a current of 33 mA, the liquid metal boils and a sharp multiple increase in its resistance occurs. It should be noted that in the solid state the bridge is capable of passing current with a density of up to $0.4\times10^{10}$ A/cm$^2$. This suggests that such a bridge is very resistant to random electrical interference that exists under real operating conditions.

This work was financially supported by the of leading Program of the National Academy of Sciences of Ukraine "Fundamental research on the most important problems of natural sciences" (section "Quantum nano-sized superconducting systems: theory, experiment, practical implementation"). State registration number of the work is 0122U001503.